\documentclass [article]{nature}
\usepackage{float}
\usepackage{units}
\usepackage{graphicx}
\usepackage{color}
\usepackage{epstopdf}
\usepackage{amsmath}
\usepackage{amsfonts}
\usepackage{amssymb}
\usepackage{dcolumn}
\usepackage{soul}
\usepackage{multirow}
\usepackage{bm}

\title{Multiple Dirac cones at the surface of the topological metal LaBi}

\author{Jayita Nayak,$^{1,\ast}$ Shu-Chun Wu,$^{1,\ast}$ Nitesh Kumar,$^1$
        Chandra Shekhar,$^1$ Sanjay Singh,$^1$ J{\"o}rg Fink,$^{1,2}$ Emile E. D. Rienks,$^{2,3}$
        Gerhard H. Fecher,$^1$ Stuart S. P. Parkin,$^5$ Binghai Yan$^{1,4~\dag}$ and Claudia Felser$^{1~\dag}$ 
         }

\date{\today}
\begin{document}
\maketitle

\begin{affiliations}
 \item Max Planck Institute for Chemical Physics of Solids, N{\"o}thnitzer Str.~40, D-01187 Dresden, Germany
 \item Leibniz Institut f{\"u}r Festk{\"o}rper- und Werkstoffforschung IFW Dresden, D-01171 Dresden, Germany
 \item Institute of Solid State Physics, Dresden University of Technology, Zellescher Weg 16, 01062 Dresden, Germany
 \item Max Planck Institute for Physics of Complex Systems, N{\"o}thnitzer Str.~38, D-01187 Dresden, Germany
 \item Max Planck Institute for Microstructure Physics, Weinberg 2, D-01620 Halle (Saale), Germany
 \\
 $^{\ast}$ These authors contributed equally to this work.

 \end{affiliations}

\begin{abstract}

The rare-earth monopnictide LaBi exhibits exotic magneto-transport properties
including an extremely large and anisotropic magnetoresistance. Experimental
evidence for topological surface states is still missing although band
inversions have been postulated to induce a topological phase in LaBi. By
employing angle-resolved photoemission spectroscopy (ARPES) in conjunction with
$ab~initio$ calculations, we have revealed the existence of surface states of
LaBi through the observation of three Dirac cones: two coexist at the corners
and one appears at the center of the Brillouin zone. The odd number of surface
Dirac cones is a direct consequence of the odd number of band inversions in the
bulk band structure, thereby proving that LaBi is a topological, compensated
semimetal, which is equivalent to a time-reversal invariant topological
insulator. Our findings provide insight into the topological surface states of
LaBi's semi-metallicity and related magneto-transport properties. 

\end{abstract}

One of the most important fingerprints of a topological state of matter is a
topological surface state (TSS).  Topological materials include topological
insulators (TIs)~\cite{Qi2011RMP,Hasan2010} and topological nodal semimetals,
that are Dirac and Weyl semi-metals~\cite{murakami2007, Young2012kz,
volovik2007quantum, Burkov2011de, Wan2011, Fang2012ga}. A topological surface
state of a TI is commonly observed as a Dirac-cone type dispersion inside an
insulating bulk energy gap~\cite{zhang2009, xia2009, Chen2009}, while a
topological surface state of a Dirac or Weyl semimetal is characterized by Fermi
arcs~\cite{Liu2014bf, Xu2015Na3Bi, Lv2015TaAs, Xu2015TaAs, Yang2015TaAs}.
However, it is challenging to identify the topological nature of surface states
for a family of gapless TIs that are characterized by the non-trivial $Z_2$ type
topological invariants, dubbed $Z_2$- topological metals ($Z_2$-TMs), due to the
lack of a bulk energy gap. For instance, Dirac-like surface states have been
found to overlap strongly with bulk states below the Fermi energy in the gapless
Heusler TI compounds~\cite{Liu2016Heusler}. Only recently, the well known Rashba
surface states of the element Au have been identified as topological surface
states~\cite{Yan2015}. The family of rare-earth monopnictides La$X$ ($X$ = P,
As, Sb, Bi) can also be classified as $Z_2$-TMs based on band structure
calculations~\cite{Zeng2015}. Moreover, a very large, unusual magnetoresistance
has been observed in LaSb~\cite{Tafti2015}, LaBi~\cite{Kumar2016} and a similar
compound YSb~\cite{Ghimire2016}, thus stimulating interest in directly observing
any topological surface states.  By contrast, angle-resolved photoelectron
spectroscopy (ARPES) on LaSb has revealed that this compound has a topologically trivial
band dispersion\cite{Zeng2016}. 

LaBi is the compound with the strongest spin-orbit coupling (SOC) in the family
of rare-earth monopnictides. In this Letter, we have investigated its
topological surface states by ARPES and $ab~initio$ calculations. Three Dirac cones have been identified in the
surface band structure, unambiguously validating the topologically non-trivial
nature of LaBi.

\section*{Results}
\begin{figure}
\includegraphics[width=0.8\textwidth]{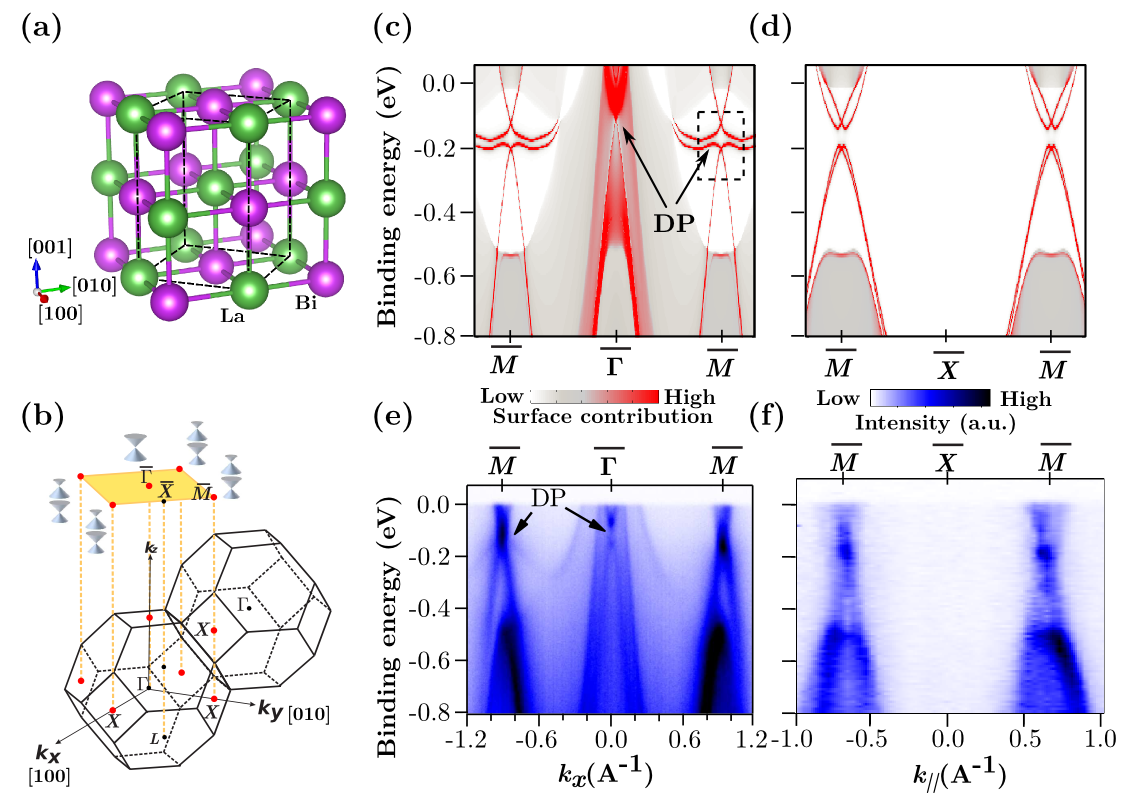}
\caption{
 {\bf Illustration of Fermi surface and the Dirac cones.}
 (a) Crystal structure of LaBi possessing simple NaCl type structure.
 (b) Bulk Brillouin zone of LaBi and the projection of the (001) surface Brillouin zone. 
 {\color{black} In bulk, the band inversion occurs at the $X$ point. Two bulk $X$ points are projected to the surface $\overline{M}$ point while one bulk $X$ point is projected to the surface $\overline{\Gamma}$ point. Thus, two Dirac-cone-like surface states are expected to exist at $\overline{M}$ and one Dirac-cone-like surface state at $\overline{\Gamma}$, as is illustrated in graph (b) showing the surface Brillouin zone.}
 (c) Calculated (001) surface band structure of LaBi along the $\overline{M}$-$\overline{\Gamma}$-$\overline{M}$ line providing the existence of a Dirac point (DP) at $\overline{\Gamma}$ and two Dirac nodes at $\overline{M}$.
(d) Calculated surface band structure along the $\overline{M}$-$\overline{X}$-$\overline{M}$ line.
(e) ARPES surface spectra measured along $\overline{M}$-$\overline{\Gamma}$-$\overline{M}$ direction with 83 eV photon energy.
(f) ARPES surface spectra measured along $\overline{M}$-$\overline{X}$-$\overline{M}$ direction with 110 eV photon energy.}
\end{figure}

LaBi crystallizes in the simple rock-salt structure (space group $Fm\overline{3}m$, No. 225), as shown in
Figure 1(a). Figure 1(b) shows the bulk Brillouin zone (BZ) and the
(001) projected 2D surface Brillouin zone of the fcc lattice. X and L points of bulk BZ are projected to
 $\overline{M}$ and $\overline{X}$ points of the surface BZ.
In the bulk band structure, the conduction and valence bands exhibit opposite
parities of wave functions and get inverted near the $X$ point~\cite{Zeng2015}.
 Although an indirect energy gap is missing with large
electron and hole pockets at the Fermi energy~\cite{Kumar2016}, the direct
energy gap still appears at every $k$-point, allowing us to define the
topological $Z_2$ invariant. We found that La-$d$ and Bi-$p$
 states contribute to the band inversion (see Supplementary Figure 1).
  Three band inversions lead to a nontrivial
$Z_2$ index $\nu_0 = 1$, which is consistent with  calculations of the parity
product of all valence bands at eight time reversal invariant $k$-points~\cite{fu2007a} that include the $\Gamma$ point, three non-equivalent $X$
points and four non-equivalent $L$ points. When these three band inversions are
projected from the bulk to the (001) surface, three Dirac-cone like surface
states appear. As show in Figure 1(b), two non-equivalent $X$ points are
projected to an $\overline{M}$ point in the surface BZ and a third $\Gamma$ point is
mapped to the surface $\overline{\Gamma}$ point. Therefore, as illustrated in
Figure 1(c), we expect that two Dirac cones coexist at $\overline{M}$, where a
direct bulk energy gap accidentally appears, and the third Dirac cone is located
at $\overline{\Gamma}$, but fully overlaps with the bulk bands.
The calculated band structures of the the LaBi (001) surface are shown in Figures 1(c-d),
 where the bright red lines represent the surface states.
The Dirac points are marked by DP. In ARPES, the Dirac point at $\overline\Gamma$ appears
150~meV below the Fermi energy whereas the surface states at $\overline{M}$ are split
into two Dirac points with energies of 123 and 198~meV below $\epsilon_F$. 
Figures 1(e-f) show the ARPES spectra  measured along the $\overline{M}-\overline{\Gamma}-\overline{M}$ and $\overline{M}$-$\overline{X}$-$\overline{M}$ directions, respectively, which are consistent with theoretical calculations.
\begin{figure}
  \includegraphics[width=1\textwidth]{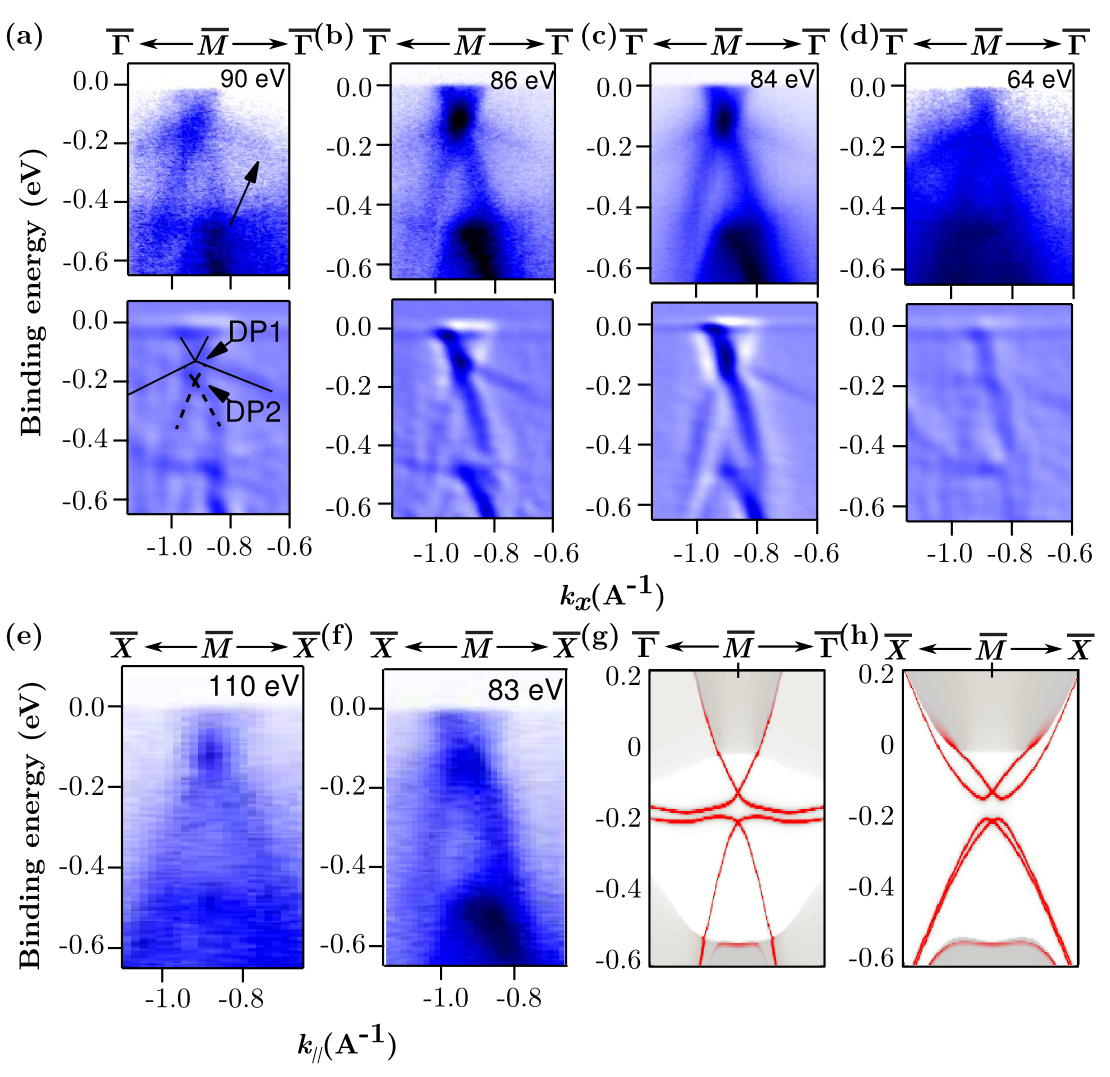}
  \caption{
           {\bf Strongly anisotropic Dirac cones at the $\overline{M}$ point.}
           (a-d) Zoomed images of the Dirac point at $\overline{M}$ at different photon energies (top panel) along  $\overline{\Gamma}-\overline{M}-\overline{\Gamma}$and its second derivative (bottom panel) to identify two Dirac points. (e,f) Dirac point at $\overline{M}$ along  $\overline{X}-\overline{M}-\overline{X}$ with photon energies 110 eV and 83 eV respectively. (g) and (h) Calculated band structure along $\overline{\Gamma}-\overline{M}-\overline{\Gamma}$ and  $\overline{X}-\overline{M}-\overline{X}$, respectively. The bright red lines indicate the surface states while the gray regions represent bulk states.
          }
\end{figure}

Details of the electronic structure around the $\overline{M}$
point have been investigated by varying photon energy and the results are
shown in Figure 2. The upper two rows show the spectra and
their $2^{\rm nd}$ derivatives along the $\overline{\Gamma}-\overline{M}-\overline{\Gamma}$ direction for
photon energies from 64 to 90~eV. 
The lower panels (e) and (f) show spectra measured along the
$\overline{X}-\overline{M}-\overline{X}$ direction for photon energies of 110 and 83~eV respectively. For
comparison, the calculated band structure is shown in panels (g) and (h) for the $\overline{\Gamma}-\overline{M}-\overline{\Gamma}$ and $\overline{X}-\overline{M}-\overline{X}$ directions, 
{\color{black} in which 
the bright red lines stand for surface states and the gray shadowed region stands for the bulk bands.
The ARPES and calculations agree quite well. 
}
The surface states are anisotropic and differ for the two directions from $\overline{M}$. The horizontal
parts of the topological surface states appear only along $\overline{\Gamma}-\overline{M}-\overline{\Gamma}$. 
The separation of the Dirac points is only about 75~meV in experiment. The splitting is
clearly resolved at 90~eV as is seen from the $2^{\rm nd}$
derivative (Figure 2(a)).{\color{black} These two Dirac-cone-like states remain at the same position in the band structure for different photon energies, indicating their surface state nature.
}
\begin{figure}
  \includegraphics[width=0.9\textwidth]{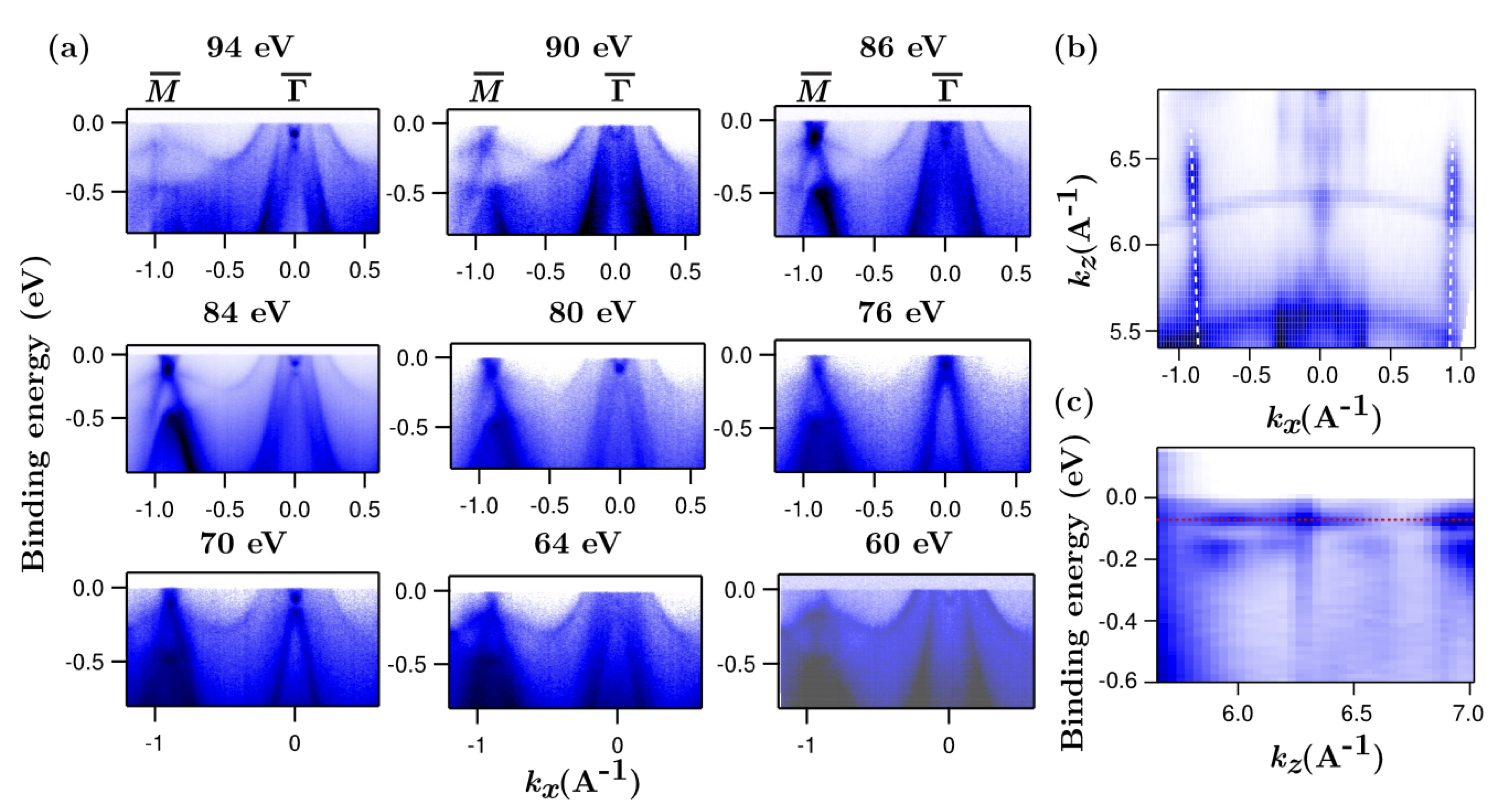}
  \caption{
           {\bf Confirmation of the topological surface state from the photon energy dependent ARPES.}
           (a) Photoemission intensity plots along $\overline{M}$-$\overline{\Gamma}$ direction at photon energies ranging from 94 eV to 60 eV with horizontal polarization.
           (b) ARPES spectral intensity maps in the k$_x$- k$_z$ plane at the binding energy of -0.12 eV, i.e., at the top Dirac point at $\overline{M}$ indicating no k$_z$ dispersion of the Dirac node (shown by white dotted lines).
           (c) $E-k$$_z$ intensity maps providing no evidence of k$_z$ dispersion of the Dirac point at $\overline{\Gamma}$ (shown by red dotted lines).
           }      
\end{figure}

To further validate the surface nature of the Dirac states, we have conducted photon energy dependent ARPES measurement.  (Figure 3).
The surface states do not disperse with photon energy (i.e. $k_z$), in contrast to the bulk states.
Photoemission spectra recorded at various
photon energies  from 60 to 94~eV (Figure 3(a)) reveal that both the Dirac cones at $\overline{\Gamma}$ and
$\overline{M}$ do not disperse with photon energies. The resulting intensity distribution
$I(k_x, k_z)$  and the dispersion $E(k_z)$ are shown in
Figures 3(b) and 3(c), respectively. The white vertical lines
in Figure 3(b) mark the intensity of the top Dirac node observed at $\overline{M}$ 
and confirms the surface nature of the bands since there is
no $k_z$ dependence throughout the whole energy range. The red horizontal line in Figure 3(c) shows the position of the Dirac point at $\Gamma$. The binding energy of the Dirac point at $\overline{\Gamma}$
does not exhibit any $k_z$ dependence, unambiguously confirming the
surface nature of the discussed Dirac bands.

We have further studied another related compound GdSb which is antiferromagnetic below 20~K. 
{\color{black} The ARPES measurement at 1 K reveals that there is no Dirac surface state at the $\overline{\Gamma}$ and $\overline{M}$ points (see Supplementary Figures 2 and 3). It is further verified by our $ab~initio$ calculations where there is no band inversions occurring in the bulk band structure (see Supplementary Figure 4). Therefore, we conclude that GdSb is topologically trivial.}
 
\begin{figure}
\includegraphics[width=1.0\textwidth]{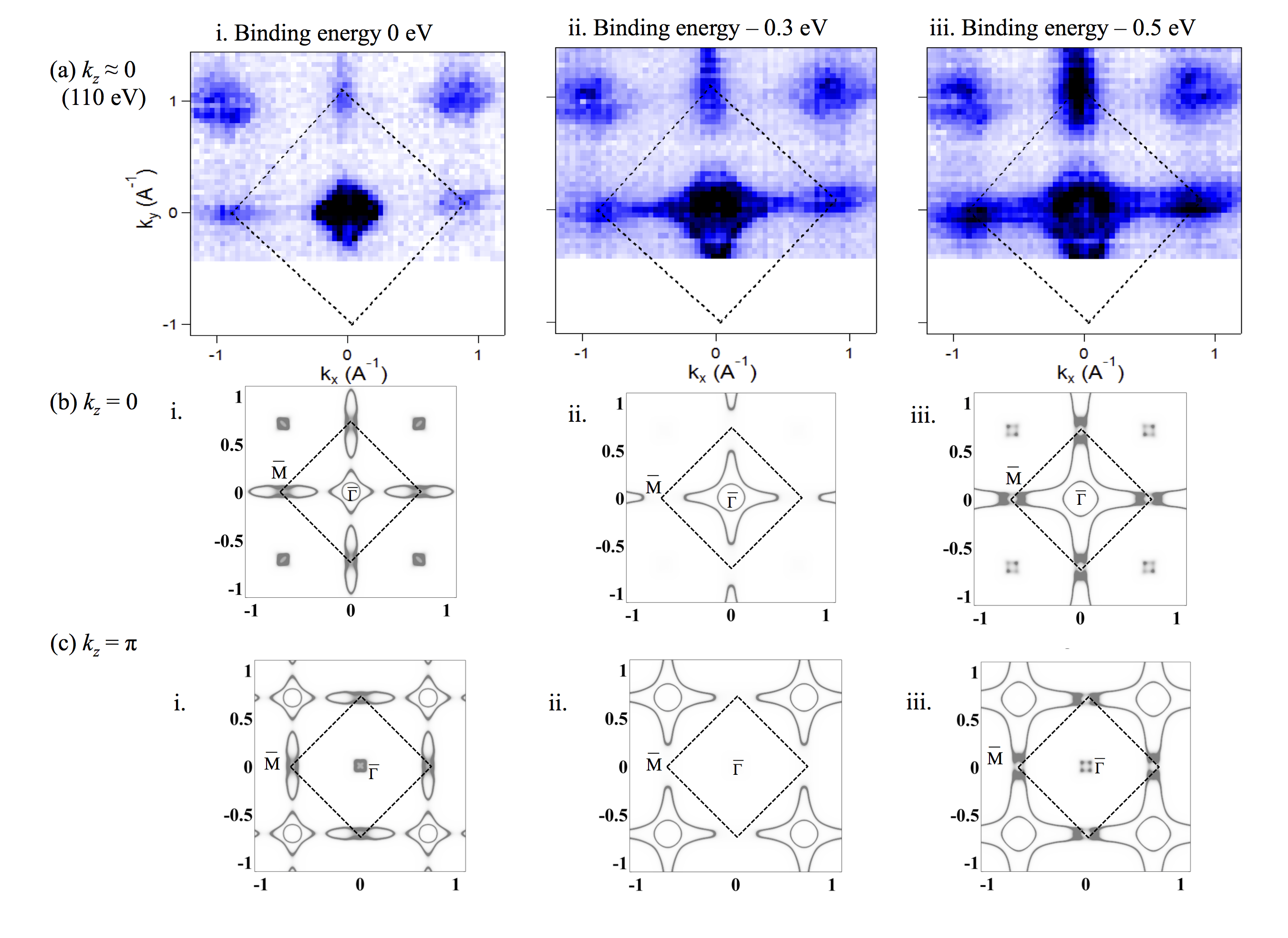}
  \caption{ {\color{black} \textbf{The Fermi surface of LaBi. (a)}, ARPES Fermi surface measured with a photon energy of 110 eV, corresponding to the bulk $k_z \approx0 $ plane.  \text{(b)-(c),} Calculated Fermi surfaces for $k_z =0 $ and $k_z = \pi$ planes, respectively. The i-iii panels stand for different Binding energies: 0, -- 0.3 and -- 0.5 eV.
          }}
\end{figure}

{\color{black}
Although the Dirac-cone-type surface states can be revealed in the ARPES band structures, signals of bulk states are still dominant in the ARPES results. It is interesting to observe the bulk bands from the Fermi surfaces. Different from surface states, the bulk bands exhibit strong $k_z$ dependence. Therefore, the Fermi surfaces look very different for different photon energies used in ARPES. According to our calculations (see Figure 4), the Fermi surface looks like a cross centered at the $\overline{\Gamma}$ point of the first BZ for $k_z = 0$. However, at $k_z = \pi$ the Fermi surface exhibits a shift to the second BZ, leaving the $\overline{\Gamma}$ point of the first BZ relatively empty. Figure 4a shows the Fermi surface measured by ARPES for $k_z \approx 0$, which is well consistent with our calculations. 
Moreover, with decreasing the binding energy, one can find that the hole pockets at $\overline{\Gamma}$ increase in size while the electron pockets at $\overline{M}$ decrease in size.
We note that the signals of surface state are too weak to be resolved in these Fermi surfaces.
} additionally, it is worth mentioning that our experimental findings about LaBi can be seen in a broader context of TIs with NaCl structure, 
where AmN and PuTe were predicted to be correlated topological insulators~\cite{Zhang2012}.

After finishing the manuscript, we began to realize recent APRES measurements on NdSb~\cite{Neupane2016} and CeBi~\cite{Alidoust2016} where only surface states at the zone corner were revealed by ARPES, 
and LaBi~\cite{Wu2016} where both zone corner and center surface bands were observed although the topological origin of these Dirac surface states was not fully recognized.

\section*{Methods}

LaBi single crystals were grown in a Bi flux~\cite{Kumar2016} and the crystal
structure was determined by x-ray diffraction using a 4 circle diffractometer.
The ARPES measurements were carried out at the UE112-PGM2b beamline of the
synchrotron radiation facility BESSY (Berlin) using the 1$^{3}$-ARPES end
station that is equiped with a Scienta R4000 energy analyzer. All measurements
were performed at a temperature of 1~K at various photon energies from 50 to
110~eV using both horizontal and vertical polarizations. The total energy
resolution was approximately 4~meV and the angular resolution was 0.2$^\circ$.
The electronic structure calculations were carried out using the local density
approximation of the density-functional theory as implemented in the Vienna {\it ab
initio} simulation package (VASP)~\cite{Kresse1993}. The generalized gradient
approximation~\cite{perdew1996} was employed for the exchange-correlation energy
functional. Spin-orbit interaction was included as pertubation. The surfaces
states were calculated by projecting the density of states to the top four atomic layers of a half-infinite surface with the Green's function method based on the tight-binding parameters from Wannier functions~\cite{Mostofi2008}.
\begin{addendum}
\item[Acknowledgements]  
This work  was financially supported  by the  the ERC (Advanced Grant No. 291472 Idea Heusler). 
\item[Correspondence] Correspondence should be addressed to B. Yan~(email: yan@cpfs.mpg.de) or C. Felser~(email: felser@cpfs.mpg.de).
\end{addendum}


\clearpage

\setcounter{figure}{0}   
\renewcommand{\thefigure}{S\arabic{figure}}
\noindent\Large\textbf{\textit{Supplemental Information}: Multiple Dirac cones at the surface of the topological metal LaBi}\normalsize

\begin{figure}[tbhp!]
\centering
\includegraphics[width=1\textwidth]{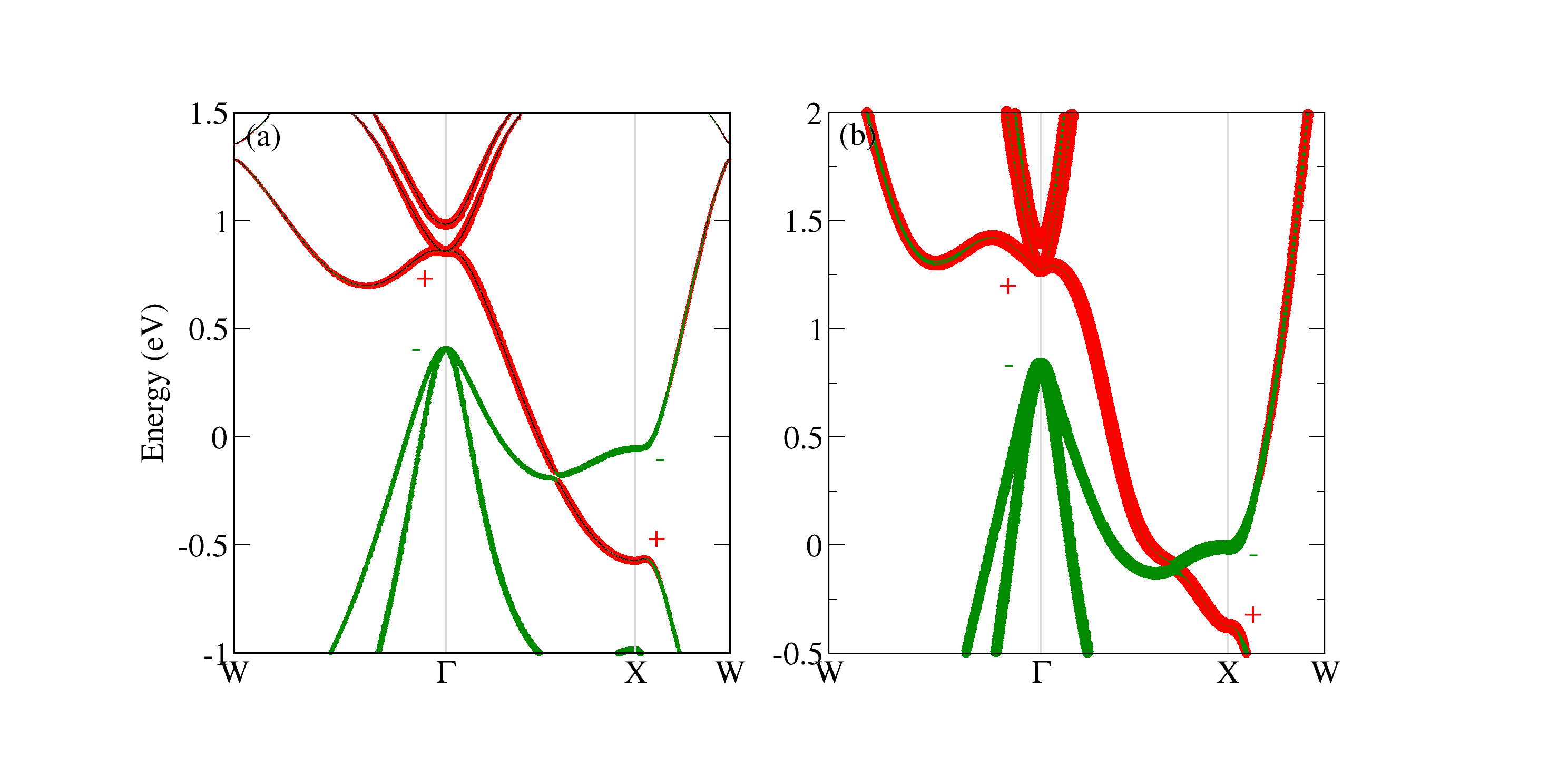}
\caption { \textbf{The bulk band structure of LaBi. (a)} Calculations within the generalized-gradient approximation (GGA) level. \textbf{(b)} Calculations in the hybrid-functional (HSE06) level. 
The Fermi energy is shifted to zero.
The red and green dots represent the orbital contribution from La-$d$ (parity ``+'') and Bi-$p$ (parity ``--'') states to the band structure, , respectively. 
It is clear that La-$d$ and Bi-$p$ bands get inverted between $\Gamma$ and $X$ points.
The parity eigen values (``+'' or ``--") are specified for the conduction and valence bands.
Since GGA is known to usually overestimate the band inversion strength, we performed HSE06 calculations that correct the GGA error, to validate the inverted band structure. 
It is clear that the topological band inversion occurs for both GGA and HSE06 calculations, resulting nontrivial $Z_2$ index $\nu_0=1$ based on the parity criteria.
}
\end{figure}
 

\begin{figure}[htb]
\includegraphics[width=0.8\textwidth]{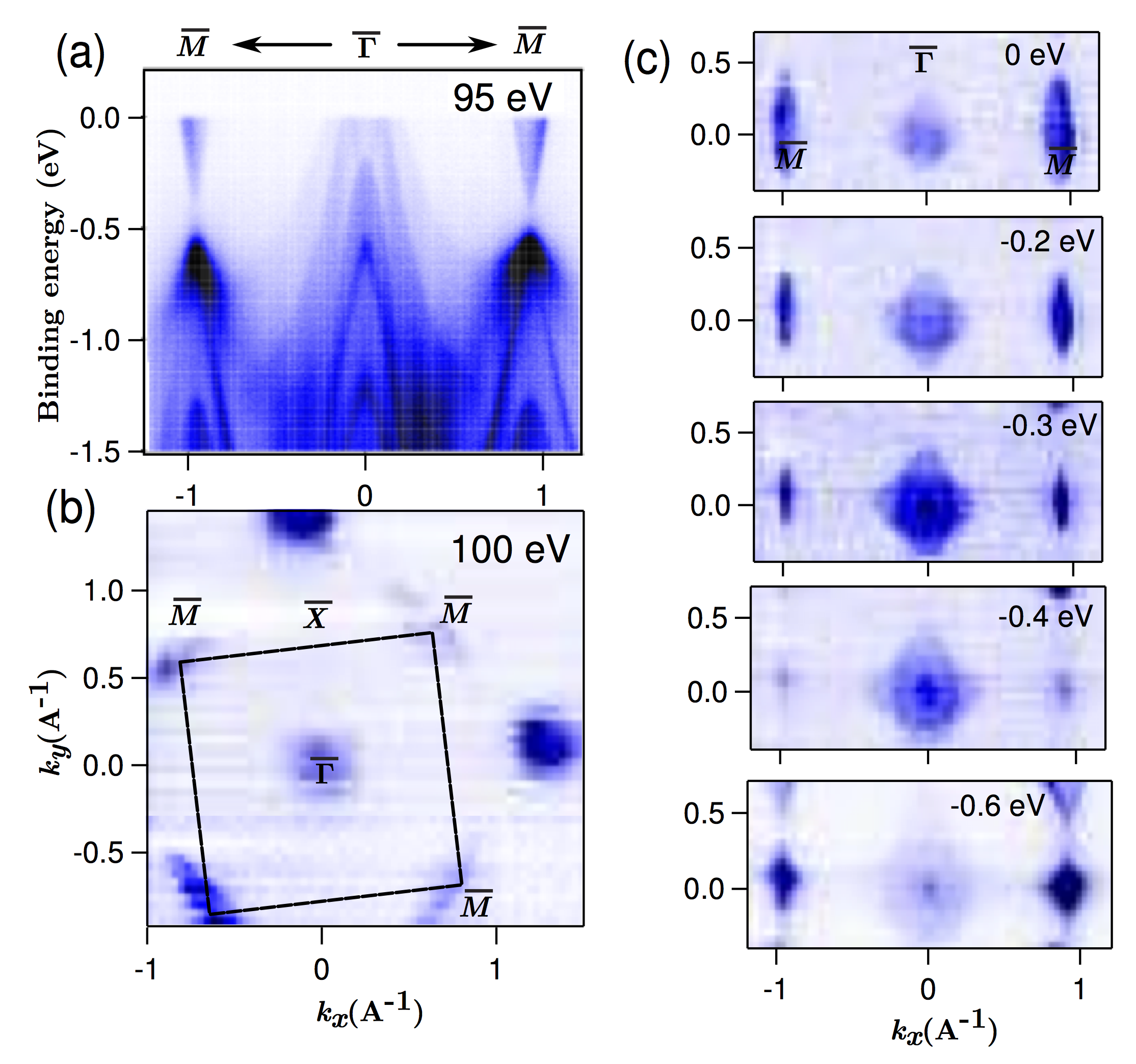}
\caption{\textbf{The ARPES measurement on the GdSb (001) surface.}
\textbf{(a)} The band dispersion along the [100] direction ($\overline{M}$-$\overline{\Gamma}$-$\overline{M}$) measured with a photon energy of 95 eV. About 0.38~eV below the Fermi energy, an energy gap of 0.16~eV is observed at $\overline{M}$ between upper and lower topologically trivial bands. There is no apparent Dirac points at $\overline{M}$ and $\overline{\Gamma}$ points. 
\textbf{(b)} The Fermi surface measured with a photon energy of 100 eV.
\textbf{(c)} The Fermi surface at different Fermi energies.
We note that GdSb which posses
the same crystal structure as LaBi, but exhibits a G-type antiferromagnetic phase below 20~K. ARPES was measured at a temperature of 1~K. From the surface states, we can conclude that GdSb is topologically trivial, distinct from LaBi.
}  
\end{figure}

\begin{figure}[htb]
\includegraphics[width=0.8\textwidth]{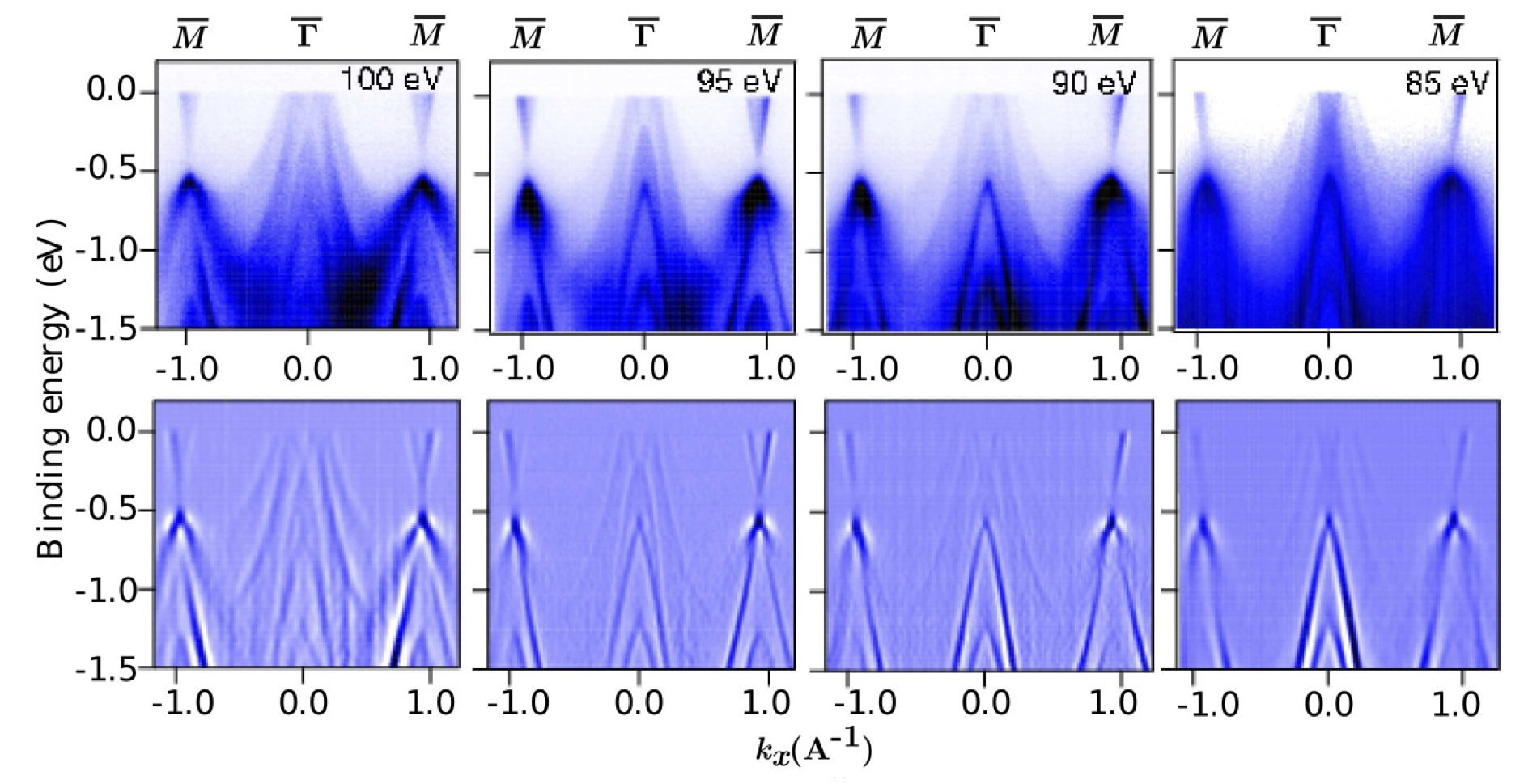}
\caption{ 
\textbf{The band structure of GdSb measured with different photon energies.}
The upper panels show the photon energy dependent ARPES spectra and 
the lower panels show corresponding second derivatives of the spectra to make details better visible.
It is clearly revealed, moreover, that no Dirac cone is detectable at
the $\overline{\Gamma}$ and $\overline{M}$ points, in contrast to LaBi(001).
        }  
\end{figure}

\begin{figure}[tbhp!]
\centering
\includegraphics[width=1\textwidth]{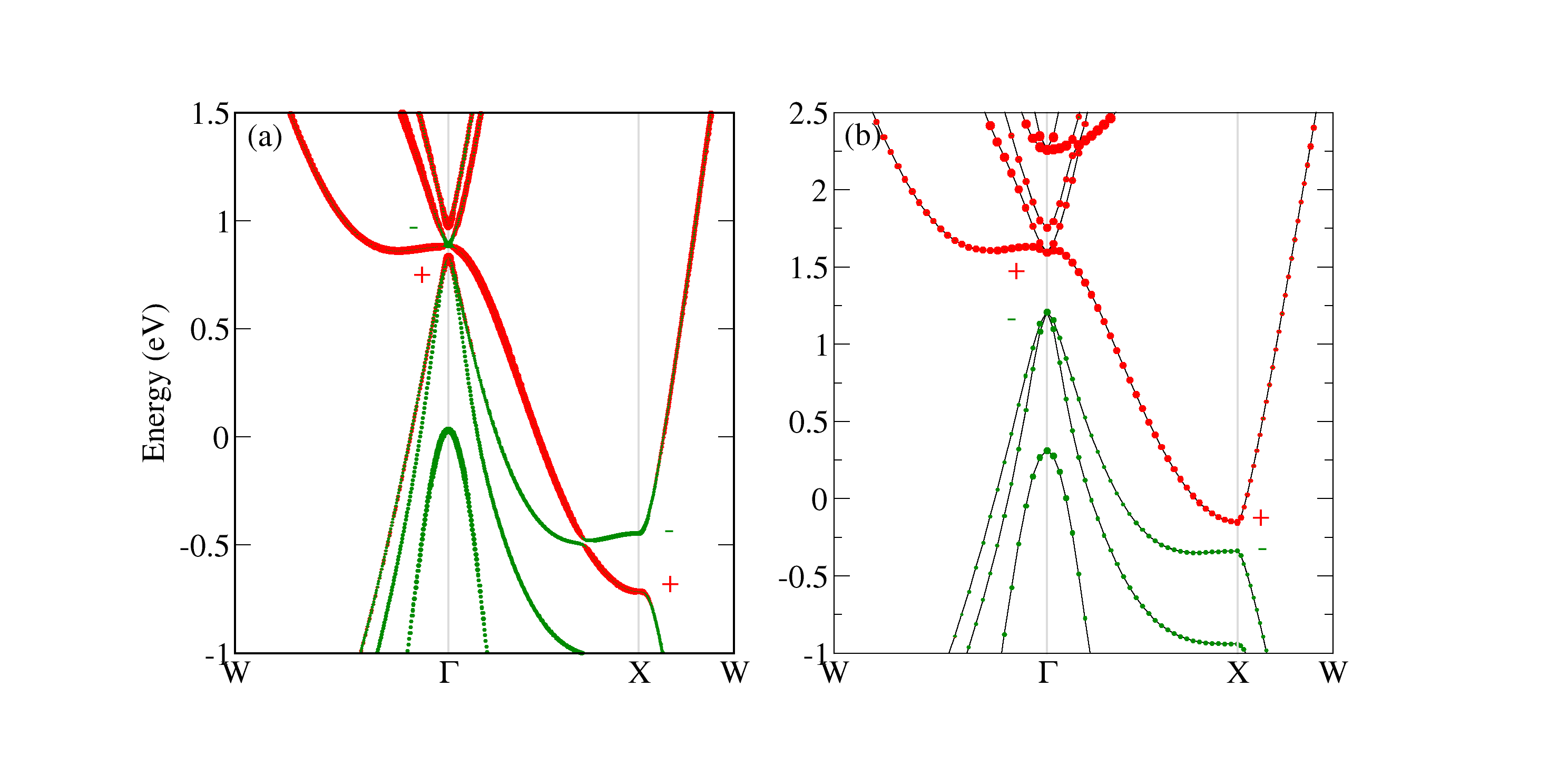}
\caption { \textbf{The bulk band structure of GdSb. (a)} Calculations within GGA. \textbf{(b)} Calculations within HSE06. The Fermi energy is shifted to zero. The Gd-$f$ electrons are frozen into the core levels in the calculation.
The red and green dots represent the orbital contribution from Gd-$d$ (parity ``+'') and Sb-$p$ (parity ``--'') states to the band structure, respectively.
In the GGA band structure, the valence band maximum (VBM) and conduction valence minimum (CBM) are Sb-$p$ states and Gd-$d$ states, respectively,  at both $\Gamma$ and $X$ points. Thus, the $Z_2$ topological invariant $\nu_0=0$ is trivial. 
In the HSE06 band structure, VBM and CBM are Gd-$d$ states and Sb-$p$ states, respectively, at both $\Gamma$ and $X$ points. Since the VBM and CBM are inverted at both $\Gamma$ and $X$, the $Z_2$ invariant $\nu_0=0$ is still trivial.
Therefore, theoretical bulk band structures indicate that GdSb is topologically trivial, which is consistent with ARPES.
}
\end{figure}

\end{document}